\documentclass[10pt]{article}
\usepackage[OE]{express}

\begin{document}
\title{Suitability and robustness of triangular nanostructured targets for proton acceleration}

\author{M. Blanco$^1$\authormark{*}, M.T. Flores-Arias$^1$\authormark{*} and M. Vranic$^2$}

\address{$^1$ Grupo Photonics4Life, Departamento de F\'{i}sica Aplicada, Facultade de F\'{i}sica, Universidade de Santiago de Compostela, Campus Vida s/n, Santiago de Compostela, E15782, Spain}
\address{$^2$ GoLP/IPFN, Instituto Superior Tecnico, Universidade de Lisboa, Lisbon, Portugal}

\email{\authormark{*}manuel.blanco.fraga@usc.es; maite.flores@usc.es} %% email address is required

\begin{abstract}
Ion acceleration in the MeV range can be routinely achieved with table-top laser technology. One of the current challenges is to improve the energy coupling from the laser to the proton beam without increasing the laser peak power. Introducing nanostructures at the front target surface was shown to be beneficial for an efficient transfer of energy to the electrons. In this manuscript, we study by using full-scale three-dimensional particle-in-cell simulations and finite laser pulses, the process when a proposed optimal target with triangular nanostructures (previously found to allow 97\% laser energy absorption) is used
. We demonstrate that the absorbed laser energy does not depend on the dimensionality in the range of parameters presented. We also present an analytical model for laser absorption that includes deviations from the ideal conditions. This is supported by a numerical parameter study that establishes the tolerance with respect to the nanostructure size, use of different ion species, existence of preplasma, etc. We found that altering the target thickness or using different ions does not affect the absorption, but it does affect the energy redistribution among the different plasma species. The optimal configuration ($h = 1~\lambda,~ w = 0.7~ \lambda$) is robust with respect to the target fabrication errors. However, high contrast laser pulses are required, because a pre-plasma layer with a thickness on the order of 0.5 lambda is enough to lower the laser absorption by more than a $10\%$ in a non-optimal scenario. 
\end{abstract}

\ocis{(190.0190) Nonlinear optics; (350.4990) Particles; (350.5400) Plasmas.}

%%%%%%%%%%%%%%%%%%%%%%% References %%%%%%%%%%%%%%%%%%%%%%%%%

%%%%%%%%%%%%%%%%%%%%%%%%%%  body  %%%%%%%%%%%%%%%%%%%%%%%%%%

\section{Introduction}\label{sec:I}
	
	Ion acceleration via laser-plasma interaction represents a promising technology for the future, and as such it has been a very active research field in the last two decades \cite{acceleration_Macchi, Daido, acceleration_several}. Ions can be accelerated efficiently in a controllable and reproducible manner into the MeV regime with table-top laser sources and metallic thin foils, via Target Normal Sheath Acceleration (TNSA) \cite{TNSA_1, TNSA_2, TNSA_mech, TNSA_proton}. 
	
	Several properties of the laser pulse and/or the target can be varied to improve different features  of the accelerated proton beam, such as the total number of particles, the maximum energy and the energy conversion efficiency. It was recently shown that nanostructuring the front target surface with a periodic pattern can lead to an improved efficiency of the laser energy absorption, and therefore increase the energy of the accelerated particles \cite{absorption_structured, absorption_structured_2, structured_1, structured_2, structured_10, structured_0, structured_3, structured_8, structured_11, structured_4, structured_5, structured_6, andreev, nanoSphere, structured_9, structured_12, structured_13, structured_14, plasmon_4, structured_15}. Theoretical models to explain this improved proton acceleration were proposed \cite{structured_1, structured_2, structured_10, structured_0} and further parameter studies revealed that the geometry of the structure is a key factor to  be taken into account \cite{structured_1, structured_2, structured_10, structured_0, structured_3, structured_8, absorption_structured, absorption_structured_2, structured_11}. Apart from ion acceleration, the nanostructured targets were also used to explore new paths for controlling high harmonic generation (HHG) and the associated attosecond pulse production \cite{structured_15, HHG_grating_1, HHG_grating_2}; to explore electron acceleration through surface plasma waves (SPW) \cite{plasmon_1, plasmon_2, plasmon_3, plasmon_4, plasmon_5, plasmon_6} or as a nanoantenna array to locally enhance the intensity of low power laser pulses \cite{nanoAntenna_1, nanoAntenna_2, nanoAntenna_3, nanoAntenna_4, nanoAntenna_5, nanoAntenna_6, nanoAntenna_7}. The variety of recent studies dealing with periodic nanostructures illustrates that their interaction with high-power table-top laser sources is of interest for several research communities.
	
	Previous numerical and theoretical studies of TNSA proton acceleration using nanostructured targets assumed ideal conditions: perfect nanostructures, no existence of pre-plasma and no laser intensity variations. Some aspects of the interaction are bound to change when one uses real targets, as the fabrication techniques have a limited precision. In addition, the laser pre-pulse generates a pre-plasma and the laser peak intensity on-target could fluctuate. Furthermore, many studies were based on 2D geometry, while it is known that removing one spatial dimension leads to an overestimation of the final proton energy \cite{2D3D, 2D3D_2}. To our knowledge, studies addressing  how deviations from ideal conditions affect the outcome of TNSA with nanostructured targets are still missing.
    
	The aim of this article is to establish how robust the experimental configurations using nanostructured targets are with respect to differences in fabrication, preplasma and laser intensity fluctuations. The final purpose is to give a more realistic estimate on the requirements of the target manufacturing and the laser contrast that both affect the final cost of obtaining the energetic ions. In a previous work published in Ref. \cite{structured_0}, some of the authours have shown through an analytical model and numerical simulations that there is an optimal configuration for laser absorption with ideal triangular nanostructured targets. In the present manuscript, we perform full-scale 3D particle-in-cell (PIC) simulations for the optimal conditions to give a plausible estimate on the accelerated proton energy. We show that by using the optimal target, ions with tens of MeV and charge on the level of nC can be obtained using table-top laser systems $(I \sim 10^{19}~$W/cm${}^2)$. We then proceed to analyze how sensitive these results are to changes in the target nanostructure and laser parameters via a series of 2D PIC simulations. We also extend our analytical model from Ref. \cite{structured_0} to predict the laser absorption in the conditions where the nanostructures on the target front surface deviate from the ideal periodic configuration. Our study reveals that the high-level of laser absorption is robust with respect to target fabrication used with the state-of-the-art techniques , but it is sensitive to preplasma and therefore laser contrast is the most restrictive requirement. 

	This paper is structured as follows: the results arising from full-scale 3D PIC simulations are presented in section \ref{sec:II}, the analysis of the effect of non-ideal conditions on the laser energy absorption is shown in section \ref{sec:III}. We consider the presence of different ion species in the target, the variation of the laser pulse peak intensity, the existence of a pre-plasma and the presence of irregularities in the periodicity of the nanostructures. Finally, conclusions are presented in section \ref{sec:IV}. All the simulations were performed with the PIC code OSIRIS \cite{osiris1, osiris2, osiris3}.
% * <marija.vranic@ist.utl.pt> 2018-04-20T12:06:03.467Z:
% 
% >  ....(list what else) 
% Add what is needed please
% 
% ^ <marija.vranic@ist.utl.pt> 2018-04-20T12:06:32.827Z.

\section{Proton acceleration via TNSA in optimal nanostructured targets} \label{sec:II}
	
	In this section we use an ideal target design to model the TNSA acceleration process for nanostructured targets with triangular nanostructures. The majority of the previous numerical studies of TNSA are performed using a two-dimensional geometry since 3D PIC modeling requires computational capabilities on the order of millions of CPU hours. Two-dimensional simulations can provide a valid qualitative description of the laser-plasma interaction. However, due to the elimination of one spatial dimension, the accelerating field at the rear surface decays at a slower rate spatially compared to a realistic 3D geometry, thus yielding to an overestimation of the proton energy. To tackle this problem and provide an estimate of what can be expected in an experiment, we perform 3D simulations for the optimal target design found in Ref. \cite{structured_0}, with a triangular structure height and width of $h=\lambda$ and $w=0.7\lambda$. Our results show that the optimal configuration retains the previously predicted enhancement in laser absorption even in a full-scale 3D geometry using a finite Gaussian laser pulse.
	
	The setup consists of a laser pulse impinging normally onto the structured surface of a solid target. The laser parameters correspond to the STELA laser at the \textit{L2A2} facility at the \textit{University of Santiago de Compostela} (USC): a Ti:Sapphire laser $(\lambda = 800~$nm$)$ with a peak power of $45~$TW and a high contrast $(> 10^{10} ~$at$~ 5 ~$ps$)$. We consider a dimensionless peak amplitude of $a_0=4$, that corresponds to an intensity of $3.46 \times 10^{19} ~$W/cm$^2$.  We use  targets with bulk thickness of $1$, $2$ and $3~\mu$m, respectively. The targets are made of electrons and heavy ions with a number density of $n = 90 n_c$, where $n_c = \frac{m_e \omega^2}{4\pi q_e^2}$ is the critical plasma density. The assumed ion mass-to-charge ratio is 9 times that of a proton, and it means to represent, on average, the mass of the different compounds that would be present in a real scenario, all with different ionization levels. Behind the ion plasma, there is a thin layer of protons and electrons with a thickness of $0.15\lambda$ \cite{TNSA_contamination_3}, where $\lambda$ is the laser wavelength. The plasma density has a steep profile. Boundary conditions are open for particles and fields in the direction of laser propagation and periodic in the transverse directions. The simulation box is $26.7\lambda$ wide and $52.5\lambda$ long. The spatial resolution is $\delta \sim 0.008\lambda$ in all directions, which corresponds to $126$ cells per wavelength. The number of particles per cell is $12$ per species. 
    
    At $t=0$, the laser pulse front is nearly touching the tip of the nanostructures. The laser has a $\sin^2$ temporal envelope, with a FWHM of $25~$fs. The transverse profile is Gaussian, focused on a $5~\mu$m spot (FWHM). The laser pulse is linearly polarized in the transverse plane, such that it is always p-polarized in relation to the nanostructures.
		
	The simulation advances in timesteps of $\sim 0.004 T$, where $T$ is the laser field period. The reflected energy is measured right after the interaction finishes ( $t = 18.75 T$), by integrating the reflected field energy density. The simulation proceeds until the accelerated protons reach a constant cutoff energy, at the time $t=238.7 T$. Energy conservation has been verified throughout all the simulations.
	
\begin{figure}[ht!]
			\centering
			\includegraphics[width=1.0\textwidth]{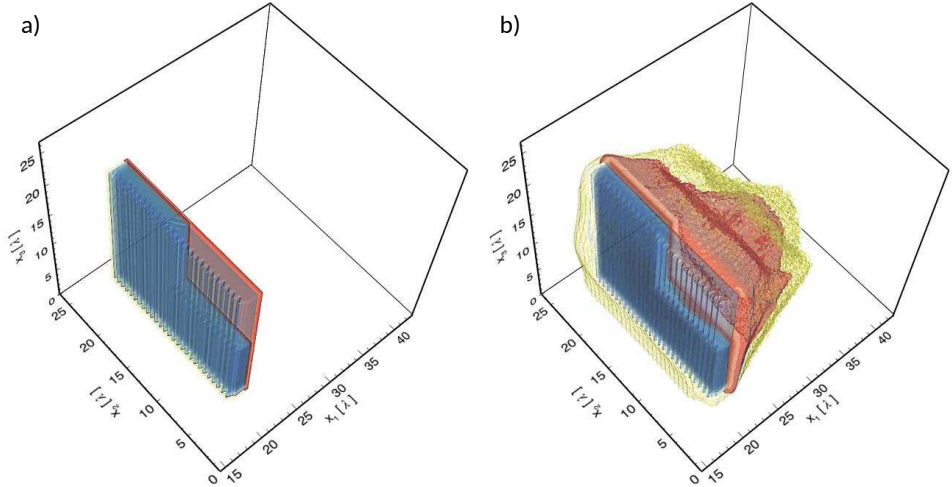}
			\caption{Scheme of the particle species distribution inside the simulation box at the a) initial instant $(t=0)$ and b) final stage of the acceleration process $(t=1200/\omega)$. The yellow isosurface corresponds to the electron density at $0.01n_c$, the blue isosurface represents ions with a density of $10n_c$ and the proton isosurface at a density of $0.02n_c$ is shown in red.}
			\label{fig3D:fig1}
	\end{figure}

	Figure \ref{fig3D:fig1} depicts the acceleration process in 3D, by showing the spatial distribution of charge density for each particle species in the target.  Panel a) shows the target with initial triangular nanostructures before the interaction, while panel b) shows the final stage of the acceleration, where the protons initially located at the rear of the target  have been accelerated.
% * <maite.flores@usc.es> 2018-06-27T08:20:22.252Z:
% 
% > $xx ~ \lambda
% 17 lambda, confirmar
% 
% ^.
	
\begin{figure}[ht!]
			\centering
			\includegraphics[width=1.0\textwidth]{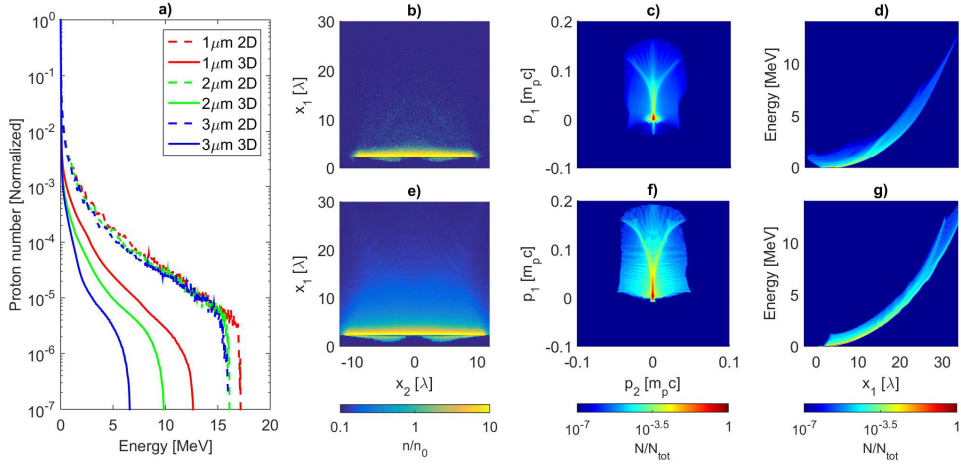}
			\caption{PIC simulations of targets with different thicknesses with an optimally nanostructured front: a) proton energy spectra for several targets;  b) ,e)  proton spatial distribution,  c) ,f) proton momentum space  $p_1-p_2$;  d),g)  proton beam energy distribution as a function of $x_1$. Panels b)-d) correspond to 3D simulations, while panels e)-g) correspond to 2D simulations.  All phasespaces b) - g) correspond to the $1~\mu$m thick target.}
			\label{fig3D:fig2}
	\end{figure}

	We have performed 3D simulations for three different target thicknesses ($1~\mu$m, $2~\mu$m, and $3~\mu$m). Figure \ref{fig3D:fig2} compares the properties of the accelerated proton beam for the 3D simulations and  their 2D counterparts. It is observed that increasing the target thickness causes a decrease in the energy of the obtained protons, being detrimental for the acceleration process. For the target with a bulk thickness of $1~\mu$m, it is possible to obtain protons with a maximum energy of $12.8~$MeV. In this case, the total charge being accelerated above $3~$MeV is $1.04~$nC, that corresponds to $6.5\cdot 10^{9}$ protons. Although the phasespaces for 2D and 3D simulations exhibit similar features, the energies achieved in the 2D case are significantly higher that those in 3D. Using a different target thickness affects the final proton energy more substantially in 3D geometry.
	
	In all the cases mentioned above (the three chosen target bulk thicknesses, 2D and 3D geometry), $\sim 97\%$ of laser energy is absorbed into the target. The agreement between the 2D and 3D results for the laser absorption is expected, as the absorption is controlled by the motion of electrons across the vacuum gaps in the laser field polarization plane. This is intrinsically a 2D problem, as the grating structures possess a translational symmetry in the third spatial direction and the relevant electron dynamics responsible for the enhanced absorption is well described in 2D. It is also not surprising that the bulk thickness does not affect the laser absorption, as changing the thickness does not change the geometry of the front surface. The thickness of the target, however, affects the redistribution of the absorbed energy among the different particle species.

	A summary of the proton beam properties for each one of the thickness target used in the example is shown in table \ref{tab3D:tab1}, where two main characteristics can be observed. On one hand, the maximum energy of the protons and the proton count above $3~$MeV, are strongly affected by the bulk thickness of the target. A decrease of $\sim 50 \%$ on the maximum energy is observed for the target $3~\mu$m thick in comparison with that of $1~\mu$m. On the other hand, the proton temperature and beam divergence are almost unchanged. This hints that the change of the bulk target thickness does not significantly alter the acceleration mechanism, but affects the efficiency by changing the way that electrons propagate through the target and distribute themselves at the target rear surface to accelerate the protons. The proton temperature is calculated by fitting the most energetic part of the spectrum to a Maxwell-Boltzmann distribution, and the beam divergence is obtained by calculating the average of  $\langle\theta\rangle = \langle\tan^{-1} (p_{\perp}/p_{\parallel})\rangle$, where the subindexes indicate the perpendicular and parallel direction with respect to the laser propagation, that is $p_{\parallel} \equiv p_1$ and $p_{\perp} \equiv (p_2^2+p_3^2)^{1/2}$. This equation is applied for the $20\%$ of the most energetic particles in the beam.
	
	\begin{table}[ht!]
		\centering
		\begin{tabular}{ | c | c | c | c | c |}
		\hline
        Thickness & Cutoff energy & Charge & Temperature & Divergence \\
		\hline
		$1$ $\mu$m & $12.83~$MeV & $1.04~$nC & $5.31~$MeV & ${6.04}^\circ$\\
		\hline
		$2$ $\mu$m & $9.95~$MeV & $0.20~$nC & $5.67~$MeV & ${6.51}^\circ$\\
		\hline
		$3$ $\mu$m & $6.72~$MeV & $0.04~$nC & $4.55~$MeV & ${7.21}^\circ$\\
		\hline
		\end{tabular}
		\caption{Properties of the accelerated proton beam for target thicknesses of figure 2a). These results are for a target with triangular nanostructures with height $h=\lambda$ and width $w = 0.7\lambda$.}
		\label{tab3D:tab1}
		\end{table}

	In summary, 3D PIC simulations show that using a table-top laser source interacting with an optimal target design,  TNSA proton beams with energies in the tens of MeV range can be obtained without increasing the laser peak intensity above $3.5 \cdot 10^{19}~$W/cm${}^2$. This is a promising result for applications such as the production of more specific tracers for Positron Emission Tomography (PET) such as ${}^{18}$F or ${}^{11}$C, that require protons with energies around $10~$MeV, according to their cross section\cite{radioisotope_1,radioisotope_2, radioisotope_3}. However, to move in this direction, one needs to consider how the acceleration changes when the conditions are less than ideal. 
	
\section{The effect of deviations from ideal conditions to the laser energy absorption}\label{sec:III}
	
The previous section demonstrated that by using an optimal structured target, one can exploit 97 \% of the laser energy for proton acceleration and obtain tens of MeV with table-top laser systems. In this section, we analyze how the deviations from the optimal configuration depicted in Ref. \cite{structured_0} affects the laser energy absorption. The optimal conditions were obtained in Ref. \cite{structured_0} using a purely electron-proton target. Here we focus on solid targets with different ion species, we vary the peak intensity of the laser pulse, consider the existence of a pre-plasma region and the effect of non-perfectly periodic nanostructures.

	\subsection{A composite target with different ion species}\label{sec:31}
    In a typical TNSA scenario the targets are manufactured from different materials that do not necessarily contain hydrogen. The accelerated protons then come from a thin layer at the target rear surface, originated from contamination with atmospheric hydrogen \cite{acceleration_Macchi, Daido}. The heavier ions from the target can also be accelerated if the contamination layer of hydrogen is removed \cite{TNSA_contamination_1, TNSA_contamination_2, TNSA_contamination_3}. Typical inexpensive metallic foils are formed of a mixture of elements that usually come as derivatives of the fabrication process, where the main element appears in a higher concentration than others. It is, therefore, important to address the effect of having a different ion species on the laser energy absorption. As the energy absorption in nanostructured targets is controlled by the electron motion within the structure vacuum gaps, we expect that the energy absorption cannot be significantly affected by changing the ion species if the ion motion is negligible on the timescale of the interaction of the laser pulse with the target. 
   
 Peak field corresponding to the laser intensity $(I \sim 10^{19}~$W/cm${}^2)$ applied without interruptions cannot displace protons further than $0.1~\mu$m from their initial position during 30 fs (which is the laser pulse duration). As the particles do not feel the peak field at all times, and the field periodically reverses its polarity, the actual displacement is even smaller. Heavier ions are slower than protons, so the ion motion begins to be significant only after the laser has been reflected or absorbed by the target.
% * <maite.flores@usc.es> 2018-06-27T08:32:20.413Z:
% 
% > TNSA mechanism starts taking place.
% Poner referencia
% 
% ^.
% * <maite.flores@usc.es> 2018-06-27T08:31:49.139Z:
% 
% > $(m_e/m_i)$
% Falta un 1/2 ?
% (I remove t)
% ^.
    
    The target design is meant to resemble that of the simulations performed in Ref. \cite{structured_0}. The targets are made of electrons and protons with a number density of $n = 40 n_c$, with a bulk thickness of $0.5 \lambda$, and  25 particles per cell per species. The density has a steep profile, as here we consider that the laser pulse has a contrast good enough to avoid the existence of a pre-plasma. The simulations are performed in slab geometry, with open boundary conditions in the longitudinal limits of the simulation box and periodic boundary conditions in the transverse direction. The dimensions of the box are: the width of $7\lambda$ and the length of $25.5\lambda$, with a spatial resolution of $\delta \sim 0.003\lambda$ in both axes, which corresponds to $314$ cells per wavelength. At the initial instant, the laser pulse is located at a distance of $0.6\lambda$ away from the target. The laser pulse propagates towards the target with a timestep of $\sim 0.0016 T$, where $T$ is the laser field period. The laser pulse has a total temporal width of 7 laser periods, with a pulse peak amplitude and polarization equal to the ones employed in the previous section. The reflected laser energy is measured right after the interaction finishes, at the time $t = 9.5 T$.
    
    We considered several different composite targets, varying the charge density and considering fully ionized ion species. Table \ref{tab:tab1} shows the absorption levels for three targets composed of electrons and different ions: hydrogen, aluminum and copper. Each configuration is considered with a plasma density of 40, 80 and $120~ n_c$, respectively. Simulation results show that the laser energy absorption is around 96 \% for all the cases, with a maximum deviation of $2 \%$. At the lowest plasma density of $40~n_c$, Hydrogen shows the largest deviation (2 \%) from the results obtained with heavier ions. However, as the density of the plasma becomes higher, the difference in the absorption reduces to values below $1\%$. According to these results, the different mixtures of ion species in composite targets are therefore not likely to change the laser absorption levels for structured targets.  
		
		\begin{table}[ht!]
		\centering
		\begin{tabular}{| c | c | c | c |}
        \hline
		Ion species $\rightarrow$	& Hydrogen & Aluminum & Copper \\
		Density $\downarrow$ & ($Z=1$) & ($Z=13$) & ($Z=29$) \\
		\hline
		$40 n_c$ & $94.8\%$ & $95.9\%$ & $96.0\%$ \\
		\hline
		$80 n_c$ & $96.1\%$ & $96.8\%$ & $96.8\%$ \\
		\hline
		$120 n_c$ & $95.9\%$ & $96.5\%$ & $96.6\%$ \\
		\hline
		\end{tabular}
		\caption{Absorption percentages for plasmas composed of electrons and different ion species at different densities. These results are for a target with triangular structures with height  $h=\lambda$ and width $w = 0.7\lambda$.}
		\label{tab:tab1}
		\end{table}
		
	\subsection{Variations in the laser intensity}
		
		A variation in the peak laser intensity can originate from fluctuations between different laser shots and laser defocusing. Due to this deviation, experiments of TNSA proton acceleration show discrepancies between different shots even in simple configurations with planar targets \cite{structured_11, fluctuation}. As typical fluctuations differ among different laser systems, we will analyze cases from small to substantial variations in laser intensity to cover a wide range of possibilities.
		
\begin{figure}[ht!]
			\centering
			\includegraphics[width=0.5\textwidth]{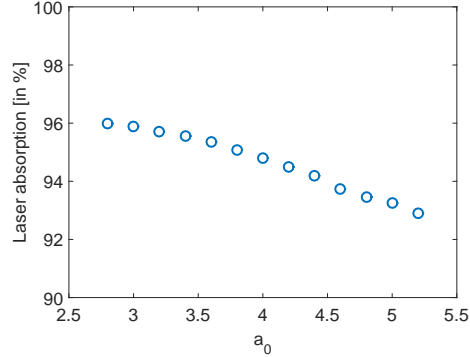}
			\caption{Laser energy absorption for different laser intensities, sweeping a $\pm 30 \%$ variation around the original laser field amplitude of $a_0=4$.}
			\label{fig:fig1}
		\end{figure}

		In the considered optimal target design where the structure height is $h=\lambda$ and width is $w=0.7\lambda$, we can observe that a $\pm30\%$ variation in the peak laser amplitude translates into a maximum $\sim 1-2 \%$ variation in the relative absorption, as can be confirmed in Fig. \ref{fig:fig1}. The simulation parameters considered are the same as in the previous subsection. This demonstrates that fluctuations in the laser intensity (in otherwise identical conditions) do not significantly alter the fraction of the transmitted laser energy. As in an experiment the total energy in a laser system is fixed, the expected value of the nanostructure-induced laser absorption enhancement is robust with respect to small fluctuations in intensity at the focal plane. 
		
	\subsection{The existence of pre-plasma}
		
        Pre-plasmas can be created through expansion of electrons and ions into vacuum, caused by the interaction with a pre-pulse or an energy pedestal before the main laser pulse hits the target. In the pre-plasma region, plasma density rises from zero to the maximum density gradually, which can result in a different interaction with a laser pulse compared to a steep density profile characteristic for the solid targets, as considered in previous sections.
		
		TNSA particle acceleration using flat targets was previously shown to be sensitive to the existence of  pre-plasmas \cite{preplasma_8, preplasma_3}. However, in the case of structured targets there are still only a few studies addressing its effect \cite{structured_3, structured_1, structured_2, structured_0}. These studies have indicated that in order to obtain high levels of laser energy absorption with nanostructures (and the increased final proton energy accordingly), the contrast of the laser pulse must be high enough to avoid the existence of a long pre-plasma that would destroy the structures before the arrival of the main laser pulse. It is not clear from previous studies whether specific features of the nanostructures can mitigate the effect of pre-plasma on the high levels of energy absorption. In this subsection we analyze how the energy absorption depends on the scale-length of the pre-plasma for different structure dimensions, and we show that some structures may be more affected than others by the existence of the preplasma. The main simulation parameters for this subsection are the same as in the previous subsection, apart from the corresponding pre-plasma.
		
		We have assumed the existence of an exponential pre-plasma profile \cite{preplasma_1, preplasma_2, preplasma_3, preplasma_4, preplasma_5, preplasma_6, preplasma_7} at the target front surface and studied its effect on the laser energy absorption for different lengths. The equation that describes the density profile is given by $n = n_0 \left[ \exp \left(\log(2) \frac{x - (x_0 - L)}{L}\right) - 1 \right]$, where $n_0$ is the plasma bulk density.  The parameter $L$, called scale-length, determines the length of the pre-plasma. This equation is defined for the spatial interval $x \in [x_0 - L,~ x_0]$, where $x$ is the spatial coordinate and $x_0$ is the position of the front of the undisturbed target. 
		
\begin{figure}[ht!]
			\centering
			\includegraphics[width=\textwidth]{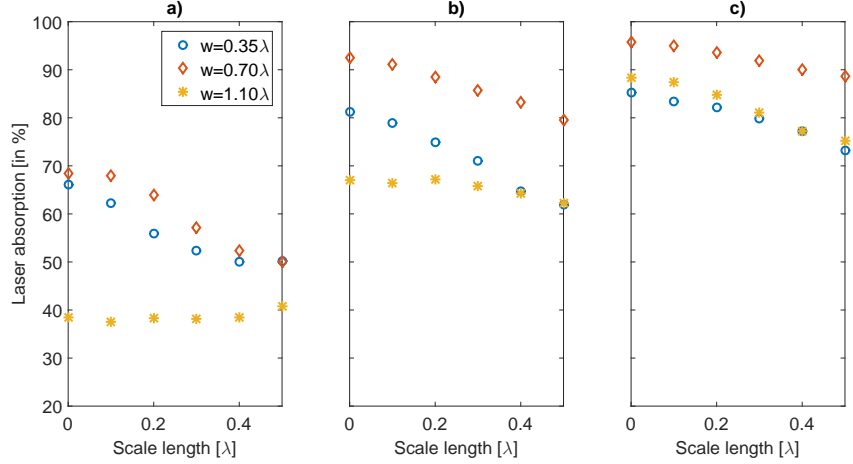}
			\caption{Effect of the pre-plasma on laser absorption. Each panel corresponds to a different structure height a) $h=0.2\lambda$, b) $h=0.5\lambda$ and c) $h=\lambda$. Diamonds correspond to the optimal structure width. Targets with smaller structures experience a higher sensitivity to the existence of pre-plasma.}
			\label{fig:fig2}
		\end{figure}

		Figure \ref{fig:fig2} shows how the presence of a short pre-plasma affects the laser energy absorption for targets with different sizes of nanostructures. For the optimized structure with height and width of $h=\lambda$ and $w=0.7\lambda$, the existence of a pre-plasma with a scale-length of $0.5\lambda$ causes a decrease of a $\sim 6 \%$ in the energy absorption. If the structure differs from the optimal, and especially if the triangles have a lower height, the effect of the pre-plasma becomes more prominent. This is evident for a structure with height  $h = 0.2\lambda$ and width $w = 0.35\lambda$, where a decrease of a $\sim 20\%$ in the absorption percentage is found for the same scale-length of $0.5\lambda$. 
% * <maite.flores@usc.es> 2018-06-27T08:35:43.753Z:
% 
% > optimized 
% optimized or optimal?
% 
% ^.
		
		The results shown in Fig. \ref{fig:fig2} demonstrate that to preserve the enhanced absorption of these structured targets, especially for a non-optimal design, very high contrast pulses are needed. Taking as a reference the expansion of pre-plasmas in flat targets \cite{preplasma_9, preplasma_10} and assuming that in these structured targets the expansion will be similar, we can conclude that a laser pulse with an ASE contrast higher than $\sim 10^{10}$, and a pre-pulse with a peak intensity not higher than $\sim 10^{16}~$Wcm${}^{-2}$ at a distance shorter than $\sim 5~$ps would be needed to have a pre-plasma shorter than $\sim 0.2\lambda$. Below this scale-length, provided that the target design is according to the optimal conditions, the effects of preplasma on the energy absorption are below a $3\%$.
		
	\subsection{Non-periodic structures}
		
		The last challenge addressed in this study is related to the non perfect periodicity of the nanostructures. This will set a requirement for the precision of its fabrication, and therefore for the fabrication costs of these targets.
		
		The fabrication of periodic nanostructured targets can be achieved by several methods, such as laser direct writing \cite{fabrication_laser_1, fabrication_laser_2}, lithographic methods \cite{fabrication_lithography_1, fabrication_lithography_2, fabrication_lithography_3, fabrication_lithography_4} or chemical etching \cite{fabrication_etching_1, fabrication_lithography_4}. The precision, velocity and costs of fabrication vary substantially depending on the chosen fabrication method and the materials employed. State of the art techniques allow the fabrication of the targets studied here with high precision. However, we aim to address what would happen if there were deviations form a perfect structure in the fabrication process, as may happen with more affordable fabrication techniques.
		
\begin{figure}[ht!]
			\centering
			\includegraphics[width=\textwidth]{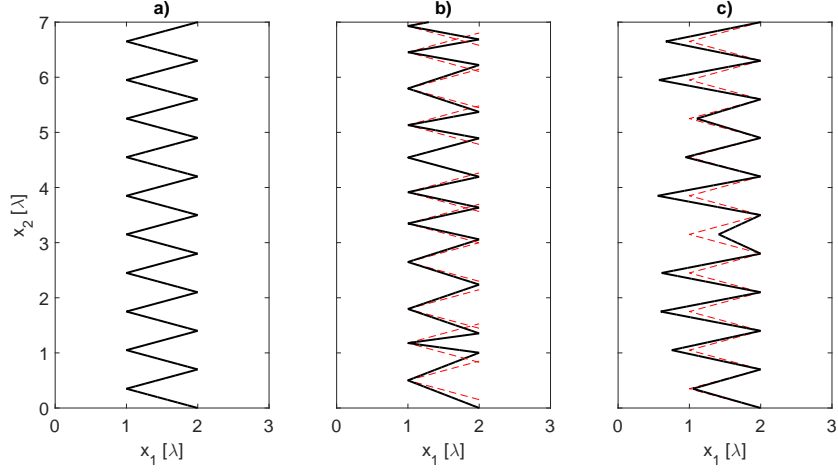}
			\caption{Examples of structured targets. a) Ideal optimal structure with height and width of $h=\lambda$ and $w = 0.7\lambda$. b) An example of a target where each individual triangle was initialized with a random deviation from the optimal value (in this example, the maximum deviation is $50\%$). c) Same as b), but the random deviation up to $50\%$ is applied to the height of each triangle. The dotted lines indicate the shape of the original optimal design of triangles centered at the same position as the ones in the current target design, to facilitate comparison.}
			\label{fig:fig3}
		\end{figure}

		To analyze the effect of irregularities in the structure periodicity, we consider targets where each triangle could, in principle, be different than its immediate neighbors. We generate such targets starting from an optimal configuration by randomly selecting the deviation in size for every individual triangle of the target. This process is controlled by a variation percentage, that sets the maximum allowed deviation from the ideal height or width for each triangle. For example with a variation percentage of $50\%$ and for an ideal structure height of $h$, all the triangles on the target would have a random height in the interval $[0.5h,~1.5h]$. Figure \ref{fig:fig3} illustrates this.
		
		Such targets are bound to have different local absorption levels than the original, optimal target. An analytical estimate for laser absorption in such a target can be obtained by extending our  model presented in Ref. \cite{structured_0}, where the equations can be tailored in order to address the local changes in geometry. The starting equations to consider are the momentum and displacement of an electron under the influence of a laser field with amplitude $a_0$:
		
		\begin{eqnarray}
			p_1 = m_e c \frac{a_0^2}{2} \sin^2(\varphi) & {~~~~} & p_2 = m_e c a_0 \sin(\varphi) \\ 
			\Delta x_1 = \lambda \frac{a_0^2}{8 \pi} \left(\varphi - \frac{\sin(2 \varphi)}{2} \right) & {~~~~} & \Delta x_2 = \lambda \frac{a_0}{\pi} \sin^2\left(\frac{\varphi}{2}\right)
		\end{eqnarray}
		
		\noindent where the labels $1$ and $2$ refer to the longitudinal and transverse dimensions, respectively, and $\varphi$ is the field phase. The relation between the initial height of an electron at the triangle surface and the oscillation phase at the position of reentry into the target is given by equation \eqref{eq:004} if the particle lands on a neighboring triangle ($\Delta x_1 \geq w / 2$ when $\Delta x_2 = h$) or by equation \eqref{eq:005} in case the electron reenters the target at the same triangle it originated from ($\Delta x_1 < w / 2$ when $\Delta x_2 = h$).
		
		\begin{eqnarray}
			h_0 = \frac{\Delta x_1}{2} + \frac{h}{w} \Delta x_2 \label{eq:004} \\ 
			\Delta x_1 = \Delta x_2 \frac{2 h}{w} \label{eq:005}
		\end{eqnarray}
		
		If we assume that the electrons carrying the laser field energy are the ones from the top of the structure ($h_0 = h$), which are the first to feel the laser field, then we solve the previous equations to obtain the phase at reentry, use this information to calculate the momentum at reentry and thus the energy they carry. By comparing that energy with a possible maximum energy attainable by the particle in a plane wave (corresponding to a reentry phase of $\varphi = \pi / 2$), one can obtain the absorbed fraction of the laser energy. This is summarized in the following equation:
		
		\begin{equation}
			A = \frac{E(\varphi)}{E(\pi / 2)} = \left(\sqrt{1 + \frac{p_1^2(\varphi) + p_2^2(\varphi)}{(m_e c^2)^2}} - 1\right) \frac{2}{a_0^2} \label{eq:006}
		\end{equation}
		
		In the case that contiguous triangles have different heights and widths, equation \eqref{eq:005} would remain the same, however equation \eqref{eq:004} would be rewritten as:
		
		\begin{equation}
			h_0 = \frac{h_A w_B}{h_B w_A} \frac{1}{\frac{h_A w_B}{h_B w_A} + 1} \left[\Delta x_1 + \frac{2 h_A}{w_A} \Delta x_2 \right] \label{eq:007}
		\end{equation}
        
		\noindent where indexes $A$ and $B$ indicate the departure and arrival triangle, respectively. It is straightforward to verify that if $h_A=h_B=h$ and $w_A=w_B=w$, equation \eqref{eq:004} is recovered.
		
		Table \ref{tab:tab2} shows a direct comparison between this model and PIC simulations, for four different targets. The ideal target design from which the random deviations were applied has triangles with height $h=\lambda$ and  width $w = 0.7\lambda$, respectively.
		
		\begin{table}[ht!]
		\centering
		\begin{tabular}{| c | c | c |}
		\hline
		Maximum variation & Model & PIC \\
		\hline
		$30 \%$ (height) & $~~94.98\%~~$ & $~~94.70\%~~$ \\
		\hline
		$30 \%$ (width) & $~~95.61\%~~$ & $~~93.89\%~~$ \\
		\hline
		$60 \%$ (height) & $~~95.03\%~~$ & $~~93.96\%~~$ \\
		\hline
		$60 \%$ (width) & $~~94.51\%~~$ & $~~94.21\%~~$ \\
		\hline
		\end{tabular}
		\caption{Laser absorption predicted by the analytical model and PIC simulations (for the exact same target) for four different examples in which each individual triangle size is randomly selected within the limits of the maximum allowed variation. The original (i.e., for ideal target before the random variation) height and width of the structures are $h=\lambda$ and $w = 0.7\lambda$, respectively.}
% * <maite.flores@usc.es> 2018-06-27T08:39:36.896Z:
% 
% > exact same 
% ???
% 
% ^.
		\label{tab:tab2}
		\end{table}
		
\begin{figure}[ht!]
			\centering
			\includegraphics[width=0.8\textwidth]{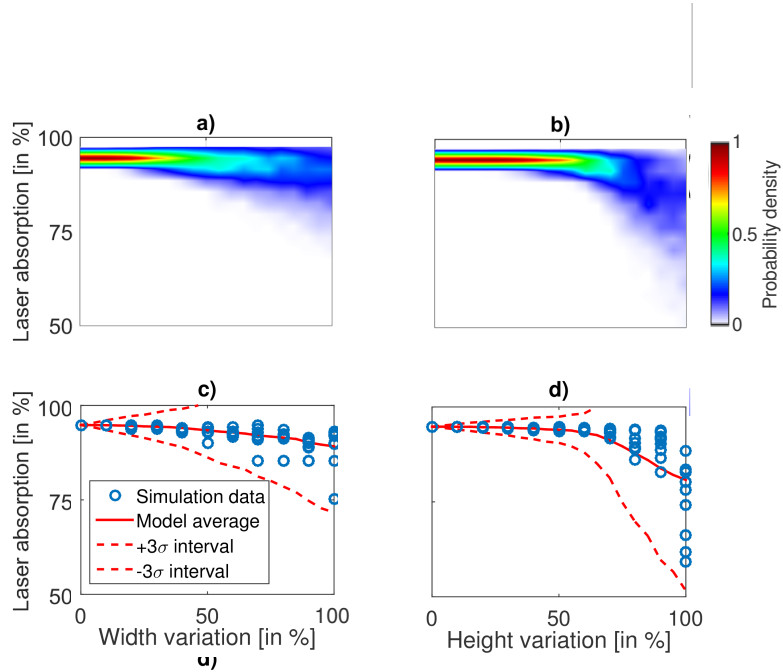}
			\caption{Results of the analytical calculations and PIC simulations for structures versus variation in height and width. The panels show the a)-b) probability density for obtaining a specific laser absorption percentage for a given maximum allowed percentage of variation of the a) width and b) height. This is accomplished by initializing 200 random configurations for each value of maximum individual size variation, and then applying the equation \eqref{eq:006} to evaluate the expected laser absorption. The c)-d) average expected absorption and $3\sigma$ intervals predicted by the analytical model are shown along with the laser energy absorption obtained from PIC simulations for different variations of the c) width and d) height.}
			\label{fig:fig4_5}
		\end{figure}

		Fig. \ref{fig:fig4_5} shows the results obtained with the analytical model taking as reference the optimized structure with dimensions $h=\lambda$ and $w=0.7\lambda$. For each variation percentage, $200$ random structures were generated and the laser absorption was computed. These individual results can be grouped to generate a probability density (panels a) and b)), that illustrates the probability for absorbing a specific fraction of the laser energy as a function of tolerance on the target fabrication (given by the maximum allowed structure size variation). The model is verified with a series of PIC simulations (main simulation parameters are given in section \ref{sec:31}) whose results are shown in Fig. \ref{fig:fig4_5}c)-d). For each value of maximum allowed variation, 10 randomly generated structured targets were considered. The results from PIC simulations are compared with the analytically predicted average and $3 \sigma$ intervals, obtained from the analytical model predictions of the probability density above. There is an agreement between the simulation data and the model predictions.

		The results shown in this section prove the robustness of these targets for energy absorption in terms of the fabrication techniques, both from PIC simulations and analytical calculations. It can be observed that even with a $50 \%$ variation in height and width, the average decrease in the energy absorption is below $10\%$. This suggests that it is possible to use cheap fabrication techniques and relax the precision requirements for the fabrication of these targets, thus reducing the potential cost of an experiment.

\section{Conclusions}\label{sec:IV}

	Triangular nanostructured targets for TNSA proton acceleration represent a very promising candidate for obtaining high energy proton sources using the table-top laser technology, as shown by the full-scale 3D simulations. An enhanced laser absorption is associated with a more efficient proton acceleration in otherwise identical conditions. The enhanced laser absorption in nanostructured targets, has proven to be very robust with respect to deviations of the topography of the structures, in terms of regularity and homogeneity from ideal fabrication.  On one hand, the effect of several factors that occur in a realistic setup, such as different ion species, variations of the laser peak intensity or defects in the regularity of the nanostructures, are found not to significantly affect the absorption for structures with height and width that we previously found to be optimal. On the other hand, it has been confirmed that in order to obtain the high absorption percentages, high contrast laser pulses must be used, especially if the target design is not optimized.  These results offer a strong motivation to use nanostructured targets in experiments and for industrial applications, as the efficiency of TNSA can be reliably increased without a notable increase in the cost of the experiment. 

	\section*{Funding}
	
	Xunta de Galicia (ED431E 2018/08) and Xunta de Galicia/FEDER (ED431B 2017/64). Spanish Ministry of Economy and Competitiveness (MINECO) (MAT2015-71119-R, FIS2015-71933-REDT). Spanish Ministry of Education, Culture and Sports (MECD) (FPU14/00289). Portuguese Foundation for Science and Technology (FCT - SFRH/BPD/119642/2016). Laserlab-Europe (EU-H2020 654148).
	
	\section*{Acknowledgments}
	
	The authors would like to acknowledge the OSIRIS Consortium, consisting of UCLA and IST (Lisbon, Portugal) for the use of OSIRIS, for providing access to the OSIRIS framework. The authors thankfully acknowledge the computer resources at \textit{Mare Nostrum 4} (RES-FI-2017-3-0029)

\end{document}